# Recent progress in parallel fabrication of individual single walled carbon nanotube devices using dielectrophoresis

Muhammad R. Islam[1, 2], and Saiful I. Khondaker[1, 2, 3] *

[1] Nanoscience Technology Center, [2] Department of Physics, [3] School of Electrical Engineering and Computer Science, University of Central Florida, Orlando, Florida 32826, USA.
* To whom correspondence should be addressed. E-mail: saiful@mail.ucf.edu

**ABSTRACT**

Single walled carbon nanotubes (SWNTs) have attracted immense research interest because of their remarkable physical and electronic properties. In particular, electronic devices fabricated using individual SWNT have shown outstanding device performance surpassing those of Si. However, for the widespread application of SWNTs based electronic devices, parallel fabrication techniques along with Complementary Metal Oxide (CMOS) compatibility are required. One technique that has the potential to integrate SWNTs at the selected position of the circuit in a parallel fashion is AC dielectrophoresis (DEP). In this paper, we review recent progress in the parallel fabrication of SWNT-based devices using DEP. The review begins with a theoretical background for the DEP and then discusses various parameters affecting DEP assembly of SWNTs. We also review the electronic transport properties of the DEP assembled devices and show that high performance devices can be fabricated using DEP. The technique for fabricating all semiconducting field effect transistor using DEP is also reviewed. Finally, we discuss the challenges and opportunities for the DEP assembly of SWNTs.

**Table of content:**

# 1. INTRODUCTION

Due to their unique electrical, mechanical, and optical properties, single-walled carbon nanotubes have attracted tremendous attention as a promising building block for future nanoelectronic devices such as field effect transistors (FET), light emitting diodes, sensors, solar cell, memory devices, [1-15] as well as electrodes for fabricating organic devices. [16-18] Although several prototypes of these devices have been demonstrated with mobility of up to 79,000 $cm^2$/Vs and transconductance of up to 6000 S/m which is significantly higher than currently prevalent silicon technology,[7, 9] a major bottleneck towards their commercial application is a reliable and cost effective technique for the scaled up fabrication of the SWNTs devices.[1, 3, 4] Scaled up fabrication of SWNT devices require assembly of SWNTs at the precise position of the circuits where the density of SWNTs can be tuned depending upon the intended application. In addition, the fabrication technique needs to be compatible with current CMOS technology which will provide added benefit of fabricating integrated circuits with SWNTs.

Scaled up integration of SWNTs were attempted via chemical vapor deposition (CVD) where nanotubes are grown directly onto the substrate using lithographically patterned catalytic island.[9, 19-23] Although CVD grown SWNTs demonstrate the best device performance including novel physics,[24-28] the high temperature (~900°C) required for the growth of SWNTs makes it incompatible with the current CMOS technologies.[29] Post synthesis assembly from the SWNTs solution is an attractive alternative to the CVD technique. Easy and simple set up, low cost, and CMOS compatible room temperature operation makes the solution-processed post growth assembly advantageous over the direct growth assembly. Several solution processed techniques for the assembly of SWNTs has been demonstrated. These includes, chemical and biological patterning,[30, 31] bubble blown films,[32] Langmuir-Blodgett assembly,[33] flow assisted alignment,[34] contact printing,[35] evaporation driven self-assembly,[36] and spin coating assisted self-assembly.[37] However, most of these techniques lack precise control over positioning and orientation of individual SWNTs. In addition, after the assembly, post etching is required to remove the excess SWNTs from the circuit. As a result, most of these techniques may not be appropriate for scaled up fabrication of individual SWNTs devices. One technique that has been effective in assembling individual SWNTs with high yield is AC dielectrophoresis (DEP).[38-40] DEP has also shown to be affective in assembling 2D and 0D materials for device applications.[41-45] DEP can be advantageous over other solution processed techniques because it allows precise control over the assembly of individual SWNTs between prefabricated electrodes at the selected positions of the circuits and does not require post-etching or transfer printing. DEP can also control the density of the nanotubes at a given site from individual to a thin film of SWNTs for various device applications.[38-40, 46-50] As a result, there is a recent surge in the research interest of the DEP assembly of SWNTs.

In this paper, we review the recent progress of DEP assembly of individual SWNTs and the electronic transport properties of the DEP assembled devices. The review begins with a discussion of the theory of DEP force and DEP assembly of SWNTs. We will present a historic overview of the DEP assembly of SWNT devices. We then discuss the role of different parameters in the DEP assembly of individual SWNT with a focus on high yield integration of SWNTs devices. This will be followed by the assembly of SWNTs from a mixed solution and semiconducting rich solution. Electronic transport properties of the devices will also be discussed. We will conclude by discussing the challenges and opportunities on the DEP assembly of SWNTs. Although this review summarizes the history and techniques for the DEP assembly of SWNTs, the most recent progress in individual SWNT devices were emphasized. For readers who are interested in network type SWNT, we refer to several recent review articles.[51-53]

## 2. Theory of DEP

Dielectrophoresis (DEP) is the phenomena wherein a polarized particle experiences a force and gains translational motion when it is placed in a non-uniform electric field.[54-56] This force was first recognized and described by Pohl in 1951.[57] When a non-uniform electric field is applied to a suspension of polarizable particles, surface charges of different polarity are induced on either side of the particle resulting in the formation of a dipole moment and cause the particle to experience a net force. As a result, the particle undergoes a translational motion and aligns along the direction of electric field lines. When the field is uniform, no net charge is induced on the particle and it does not experience any force.

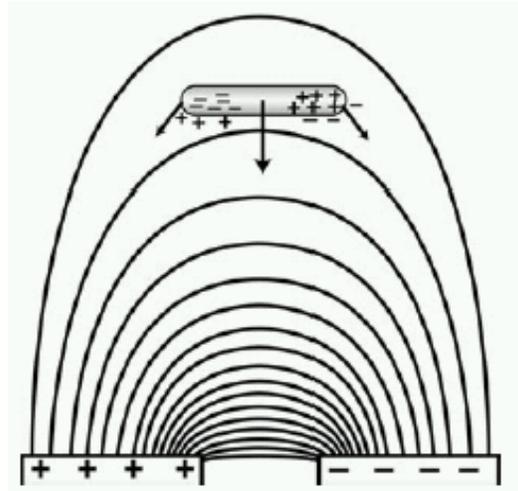

**Fig. 1.** Sketch illustrating a carbon nanotube subjected to dielectrophoresis. The nanotube immediately aligns with the field lines and experiences a net force from the ac field. This force may be directed towards the high field region (down) in the case of positive dielectrophoresis (shown), and towards the low field region (up) for negative dielectrophoresis (not shown). Reproduced with permission from [58], M. Dimaki et al., *Nanotechnology* **15**, 1095 (**2004**). © 2004, IOP Publishing Ltd.

When placed in an inhomogeneous electric field, the DEP force experienced by a polarized particle is given by $\vec{F} = (\vec{p} \cdot \vec{\nabla})\vec{E}$, where $\vec{p} = v\alpha\vec{E}$ is the induced dipole moment of the particle, $v$ is the volume of the particle, $\alpha$ is complex effective polarizability, and $E$ is the applied electric field. For a SWNT, the DEP force can be calculated from an effective dipole approximation by considering it as a cylinder. A simplified diagram of the DEP force on a SWNT is shown in figure 1. The DEP force acting on the tube with its major axis parallel to the applied electric field is given by[58,59] (new ref) $F = \frac{\pi r^2 l}{2} \varepsilon_m^* Re(K_A) \nabla |E|^2$ where $\varepsilon_m$ is the real part of permittivity of the suspending medium, $K_A$ is the Clausias-Mossotti (CM) factor which depends on both the complex dielectric constant of the medium ($\varepsilon_m^*$) and the particle ($\varepsilon_p^*$) and is given by $K_A = \frac{\varepsilon_p^* - \varepsilon_m^*}{3\varepsilon_m^*}$ where $\varepsilon^* = \varepsilon - i\frac{\sigma}{\omega}$. Here $\varepsilon$ is the real permittivity, $\sigma$ is the conductivity, and $\omega = 2\pi f$ is the frequency of the electric field. A more detailed calculation and finite element simulation can be found in ref [60].

DEP force depends on the physical properties of the SWNTs as well as the properties of the medium in which the SWNTs are suspended. Depending on the permittivity of the SWNT ($\varepsilon_p^*$) and the suspending medium ($\varepsilon_m^*$), DEP force can be either positive or negative. If $\varepsilon_p^* > \varepsilon_m^*$ then $K_A$ is positive, in this case the dipole moment will align along the electric field and is called positive DEP. The opposite case will occur if $\varepsilon_p^* < \varepsilon_m^*$ in which case the particle will be directed away from the electric field and is called negative DEP. Metallic SWNT (m-SWNT) and semiconducting SWNT (s-SWNT) experiences different DEP forces due to their different dielectric constant. Typically, m-SWNTs has a dielectric constant much larger than that of the suspending medium which cause them to experience greater DEP force compared to semiconducting SWNTs[58]. In addition, the DEP force also depends on the size of the particle, therefore particles with larger volume experience a greater DEP force. The effect of DEP parameters on the directed assembly of SWNTs will be discussed in great details in section 5.

## 3. HISTORY OF THE DEP ASSEMBLY OF SWNT

Soon after the discovery of the carbon nanotubes, DEP was used to assemble nanotubes between the prefabricated electrodes. In 1996 Yamamoto et al. demonstrated the DEP assembly of multi walled nanotubes (MWNTs).[61,62] Thin films of MWNTs bundles were assembled from a solution containing a mixture of MWNTs and carbon particles (Fig. 4(a)). DEP assisted alignment of SWNTs was first reported by Chen et al. in 2001.[63] A thin film of SWNTs bundles (Fig. 4(b)) were assembled by immersing a substrate containing source and drain electrodes in SWNTs solution and then applying AC electric field between the electrodes. It was shown that the density and alignment of SWNTs bundles depends on the frequency and magnitude of applied AC voltage. Since the SWNTs have strong van der walls interaction, they prefer to remain in bundle.[64] If a solution contains a lot of bundles, the DEP will assemble the bundles because of their high dielectric constant together with larger volume. Besides, the uniform electric field produced between the planer electrodes used in this study resulted in the assembly of thin films.

Assembly of single SWNT bundles was first demonstrated in 2003 by Krupke et. al [65, 66] (Fig. 4(c)). In addition, taper shaped electrode was used which helps assembly of single bundles as the field was strongest at the edge of the tip. Unlike previous studies, they used purified SWNTs suspended in *N, N*-dimethylformamide (DMF) solvent. However, the absence of any stabilizer re-aggregates the SWNTs into bundle resulting in the assembly of individual bundles due to their high dielectric constant and larger volume. Since the bundles contain metallic SWNTs, the assembled devices show metallic behavior. However, these authors demonstrated that FET characteristics can be achieved from such a device after selectively breaking the metallic SWNTs in the bundle shown in Fig. 4(d). Controlled assembly of bundle of SWNTs by varying different DEP parameters was also demonstrated by several research groups.[67-69]

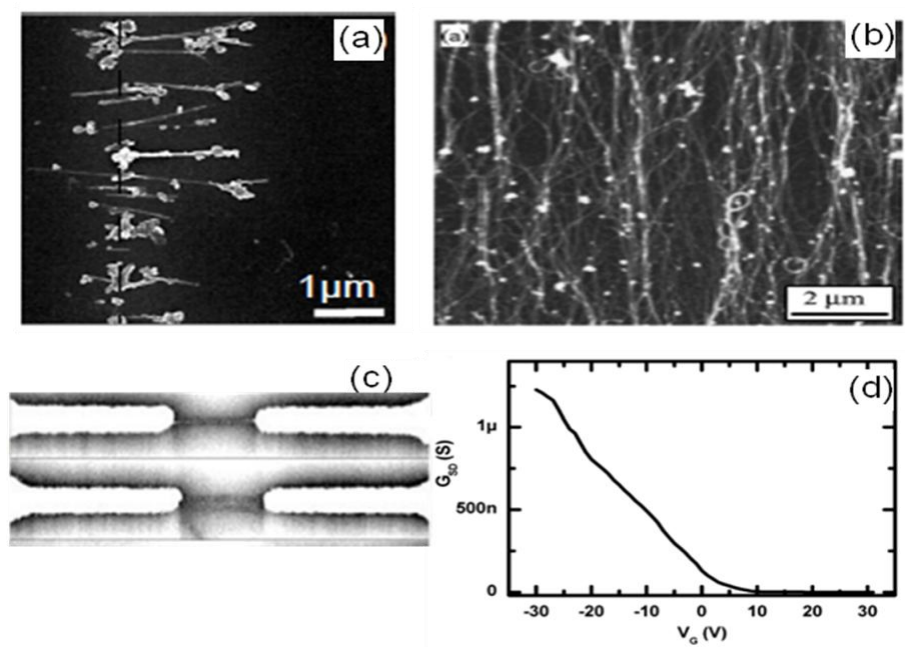

**Fig. 4.** (a) SEM images of a thin film of bundle of MWNT aligned by the application of electric field. Reproduced with permission from [62], K. Yamamoto et al., *J. Phys. D-Appl. Phys.* 31, L34, (**1998**). © 1998, IOP Publishing Ltd. (b) Tapping-mode AFM images of a thin film of bundle of SWNT aligned by applying DEP. Reproduced with permission from [63], X. Q. Chen et al., *Appl. Phys. Lett.* 78, 3714 (**2001**). © 2001, American Institute of Physics. (c) SEM image of single bundle of SWNT trapped by DEP, Separation between the electrodes are 400 nm. (d) Source-drain conductance $G_{SD}$ vs gate voltage $V_G$ for a single bundle of SWNT formed by ac-dielectrophoresis giving p-type FET behavior after selective burning of metallic SWNT by high voltage pulse. Reproduced with permission from [65], R. Krupke et al., *Nano Lett.* 3, 1019 (**2003**). © 2003, American Chemical Society.

For the assembly of individual SWNTs, a bundle free, individually dispersed SWNTs solution is needed. Usually this can be achieved either by noncovalent adsorption or covalent functionalization of SWNTs. In noncovalent adsorption, SWNTs are encapsulated with surfactants by long and aggressive sonication times. While in covalent functionalization, carboxylic groups are introduced to the SWNTs sidewall through acid treatments which separate the bundles and stabilize the suspension.

It was shown by O'Connell et al. that individually dispersed SWNTs solution can be obtained by immersing the as grown SWNTs in a aqueous solution containing Sodium dodecyl sulfate (SDS).[70] Advent of highly effective surfactant for individually dispersed SWNTs [70-75] facilitates the DEP assembly of individual SWNTs. Site selective DEP assembly of individual metallic and semiconducting SWNTs was demonstrated by Zhang et al. in 2005[76,77] using a surfactant stabilized SWNTs solution (Fig. 5(a)). Around the same time, controlled and reproducible alignment of individual SWNTs was demonstrated using floating-potential DEP method.[64] Individual SWNTs assembled by these techniques contain both metallic and semiconducting SWNTs.[76-80] The average assembly yield in these studies were ~ 26%.

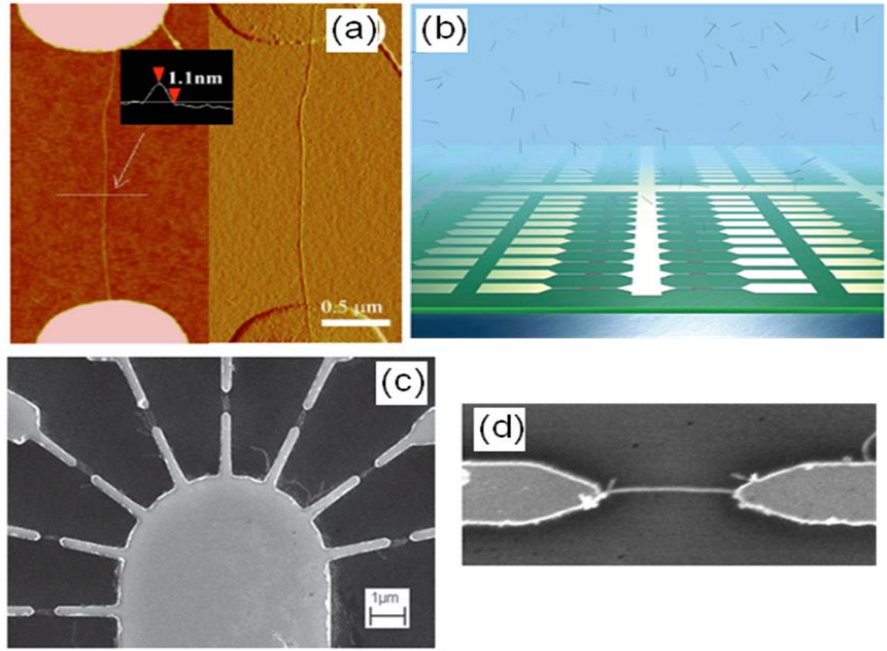

**Fig. 5.** (a) AFM images images of of an individual semiconducting SWNT with diameter 1.1 nm deposited by DEP. Reproduced with permission from [77], Z. B. Zhang et al., *J. Appl. Phys*. 98, 056103 (**2005**). © 2005, American Institute of Physics. (b) Schematic of the high-density array of single-tube devices, comprising interconnected biased electrodes and counter electrodes capacitively coupled to the p-type silicon substrate $SiO_2$. Reproduced with permission from [38], A. Vijayaraghavan et al., *Nano Lett*. 7, 1556 (**2007**). © 2007, American Chemical Society. (c) SEM image of high-density device array fabricate using semiconducting rich (99%) SWNT solution. Reproduced with permission from [91], M. Ganzhorn et al., *Adv. Mater*. 23, 1374 (**2011**). © 2011, John Wiley & Sons, Inc. (d) SEM image of individual SWNT assembled between source-drain contact with a separation of 1μm via DEP using all s-SWNT solution. Reproduced with permission from [92], M. R. Islam et al., *Nanotechnology* 23, 125201 (**2012**). © 2012, IOP Publishing Ltd.

Ultra-large-scale assembly of individual SWNT was demonstrated by Vijayaraghavan et al. in 2007 shown in Fig. 5(b).[38] The simultaneous deposition technique with capacitively coupled counter electrodes was used to assemble a high density of individually contacted SWNTs within a short period of time. By using a thick oxide layer of 800 nm, the authors obtained a 90% assembly yield for individual SWNTs. However, the thick gate oxide resulted in higher subthreshold swing together with large hysteresis.

Over the last few years, there has been tremendous research effort and continuous progress in producing high quality, individually dispersed, and stable SWNTs solutions.[81,82] This has led to the commercialization of not only high quality mixed SWNTs solution but also chirally sorted metallic and semiconducting SWNT solutions.[83,84] Controlled assembly of individual and array of SWNT devices using high quality SWNTs solution was demonstrated by Khondaker group.[40, 46, 47, 85, 86] The effect of various DEP parameters on the integration of SWNTs and performance of the assembled devices were reported in their works. They showed that individual SWNTs devices made from commercially obtained mixed SWNTs solution presents improved device performances and can be comparable to those of CVD grown SWNTs [40, 60]. It was also demonstrated that the SWNTs contains much less defects or no defect which gives rise to high quality device performance.[87]

One of the major bottleneck in DEP assembly is that when a mixed solution of SWNTs is used, a semiconducting device yield of 50% or lower[39, 40, 77, 78] is obtained as metallic SWNTs feel greater force over semiconducting during the DEP assembly. Since only s-SWNTs can be used for FET application, using a mixed solution leaves a large percentage of devices non-functioning. Recently, solution based SWNTs sorting techniques have been used to separate nanotubes by chirality.[37, 88-90] Chirally shorted all semiconducting SWNTs (s-SWNTs) solution had been used to assemble s-SWNTs by Ganzhorn et al.[91], and Islam et al.[92] (shown in Fig. 5(c),and 5(d) respectively). A significant improvement in the FET yield (up to 99%) had been demonstrated using a semiconducting-enriched, high quality s-SWNTs aqueous solution.[91-94]

## 4. Experimental SET up for the DEP assembly of individual SWNT

For the assembly of individual SWNTs via DEP, pre-fabricated metal electrode patterns on a suitable substrate is required. Since the assembly is done in an ambient environment and the SWNTs are solution processed, the DEP assembly technique is compatible with any substrate including plastic. Figure 2 (a) shows a cartoon of a typical DEP assembly set up. Here, the metal (Pd) electrode patterns are fabricated on a Si/SiO$_2$ substrate. The highly doped Si can be used as a global back gate while the thermally grown oxide works as a gate insulator. The metal electrode pattern has a common source for simultaneous deposition of many SWNTs between the common source and individually accessible drain electrodes. The electrode patterns can be defined by electron beam lithography (EBL) or optical lithography depending on the gap size between the electrodes. If the gap size is larger than 1 µm, optical lithography can be used while if the gap size is 1 µm or less, typically EBL is used to define the pattern. The design of the electrode gap depends on the length of the SWNTs in the solution. Although as grown SWNTs can be tens of microns long, however, solution processing of the SWNTs breaks them to a few micron length. Since, the aim is to connect individual SWNTs between the source and drain electrode patterns, the typical electrode gaps are 1-2 µm to match the median length of SWNTs in the solution. After defining the pattern through lithography, a thin film of metals is deposited by thermal or electron beam evaporation followed by lift off. Pd or Au is widely used for the fabrication of SWNTs based field effect transistors as these materials give best contact with SWNTs.[95-97]

The next step in the DEP assembly involves casting of a small drop of SWNT solution of optimized concentration onto the chip containing the electrodes. An AC voltage of desired frequency is then applied between the common source electrode (SE) and gate electrode (GE) while floating the drain electrodes (DE). The AC voltage is applied for a short period of time (~1 to 3 min) depending on the solution concentration. Since the DEs are capacitively connected to the GE, they experience a similar potential as that of GE. As a result, a potential difference exists between each SE and DE which allows parallel integration of a large number SWNTs between the electrode gaps within a short time. This is called simultaneous deposition technique as it allows the deposition of a large number of SWNTs simultaneously by applying the AC voltage only once. It is expected that, as soon as one SWNTs bridges a SE - DE electrode pair, the potential difference between the electrodes are significantly reduced which would prevent further deposition of SWNTs in the channel. Figure 2(a) show a schematic of a Pd electrode pattern fabricated on Si/SiO$_2$ substrate[40]. One important thing to notice is that the electrodes are taper shaped. Simulated image of the electric field created by the taper shaped electrodes due to AC voltage are shown in Fig. 2 (b).[85] It can be seen that for a taper shaped electrode the electric field is strongest at the tips, which in turn increase the probability of the alignment of individual SWNTs between the tips. If the electrodes are not taper shaped and have considerable width, it can be challenging to optimize the assembly of individual SWNTs as those electrode patterns can attract many SWNTs simultaneously. Even though the use of common back gate increases the DEP assembly yield, however using back gate for transistor operation can lead to a lower switching speed which can be circumvented using local gates discussed in section 6.

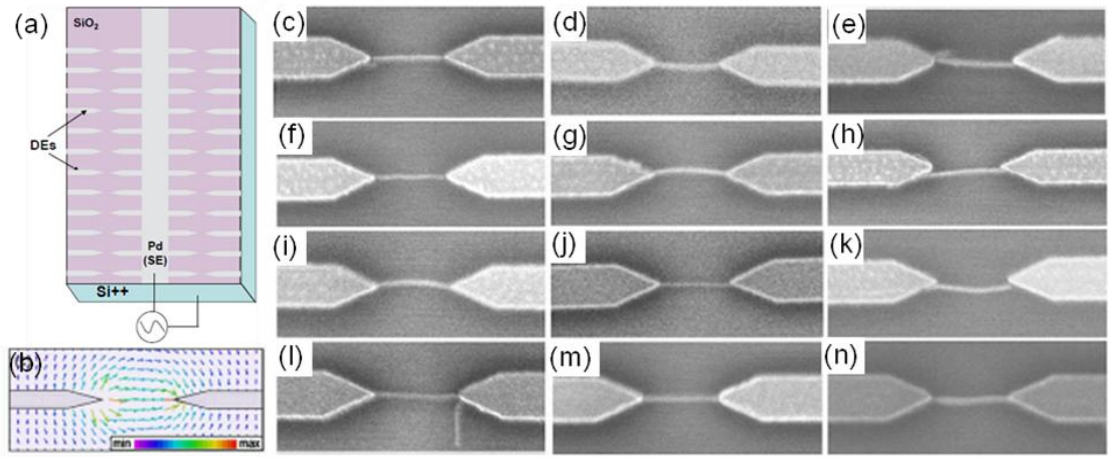

**Fig. 2.** (a) Schematic of the electrode array and DEP assembly setup for simultaneous deposition. Reproduced with permission from [40], P. Stokes et al., *Appl. Phys. Lett.* 96, 083110 (**2010**). © 2010, American Institute of Physics. (b) 2D simulated electric field around the electrode gap for the taper shaped electrodes and (c)-(n) SEM images of several individual SWNT assembled via DEP from commercial solution at a concentration of ~10 ng/ml. Gap between the electrodes is 1 μm. Reproduced with permission from [85], P. Stokes et al., *J. Vac. Sci. Technol. B* 28, C6B7 (**2010**). © 2010, American Institute of Physics.

DEP assembly of individual SWNTs can also be achieved by using individual source and drain electrode (without a common source). In this case, a series of parallel SE and DE's are fabricated and an ac voltage is applied between each pair of SE and DE and then moved to the next pair. The schematic of this assembly is shown in Fig. 3 (a). Here, the time taken to assemble individual SWNTs between each pair can only be a few seconds.

Sometimes an external resistor (~ GΩ), called a self-limiting resistor is used in series with the DEP set up.[98, 99] This is to ensure only an individual SWNTs is assembled between the electrodes. When an individual SWNTs is connected between the electrodes, the resistance between the electrode gap becomes much smaller (~MΩ) than the limiting resistor and the entire applied voltage drops across the limiting resistor. In other words electric field between the electrode gap is reduced which prevents further deposition of SWNTs between the electrodes.

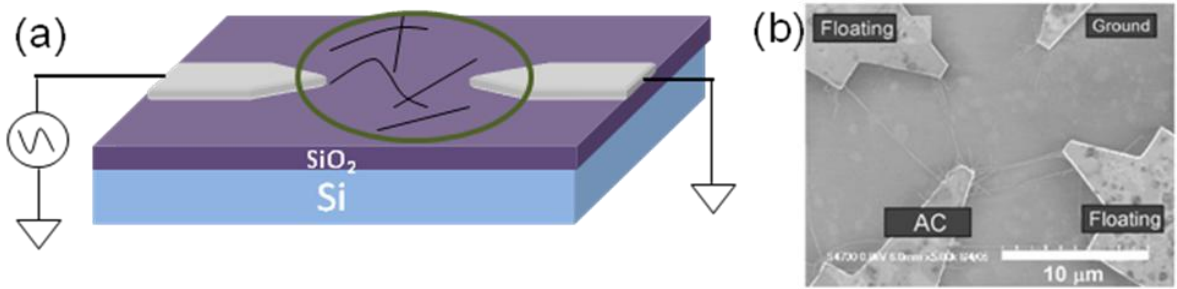

**Fig. 3.** (a) DEP assembly setup for the assembly of SWNT using individual source and drain (b) SEM image shows a single nanotube guided along the edge of 1μm-wide strips using floating potential technique. Reproduced with permission from [79], Benerjee et al., *J. Vac. Sci. Technol. B* 24, 3173 (**2006**). © 2006, American Institute of Physics.

Another technique that has been used for DEP assembly of SWNT is called floating potential technique.[78, 79] This is shown in Fig 3 (b). In floating potential method, there is a SE where the electric field is applied (often called control electrode) while several additional metal posts are placed nearby. This technique is similar to simultaneous deposition technique where the floating posts get the potential due to capacitive coupling with the source. However, in this technique the assembled SWNTs need to be connected to drain electrode with additional lithography step.[78, 80]

Following the DEP assembly, the deposition yield is typically checked via Scanning electron microscopy (SEM) or atomic force microscopy (AFM). The quality of the assembly depends on the quality of the solution. If there are catalytic particles in the solution, the particles will also be assembled along with the SWNTs. In addition, if there are bundles of SWNTs, they will be preferentially assembled as the DEP force is greater for the bundle.[65, 85] Figure 2(c)-(n) show SEM images of a number of individual SWNTs assembled by DEP from a catalytic particle free, high quality SWNT solution.[85] It can be seen here that the devices are free from bundles or any additional short SWNTs in the channel. Typical assembly yield for individual SWNT varies between 20% - 30%.[38-40, 78] In one report, an assembly yield of up to 90% was reported.[38] The assembly of individual SWNT and its yield depends on the nature and quality of the solution, solution concentration, applied frequency, voltage applied, DEP time, shape of electrodes, and substrate oxide thickness.[56, 59, 60, 92, 100-109] The effects of those parameters on the assembly of SWNTs will be discussed in the section 5.

## 5. PARAMETERS CONTROLLING THE DEP ASSEMBLY OF INDIVIDUAL SWNT

Assembly of SWNTs via DEP depends on several parameters such as the quality of the SWNTs solution, the magnitude of the applied AC voltage, frequency of the AC signal, size and shape of the electrodes, deposition time and substrate oxide thickness. Since DEP is a solution processed assembly technique, the assembly of individual SWNT and their yield are greatly influenced by the quality and stability of the SWNT solution. For high yield assembly of SWNTs, the solution needs to be catalytic particles free, stable, and the nanotubes need to be uniformly dispersed. Being catalytic particle free is important, as catalytic particles in the solution tend to make their way into the electrode gap along with the SWNTs during the assembly process. The catalytic particles are highly conductive and their presence in the device can disrupt the device performance. The effect of catalytic particle on the DEP assembly of individual SWNT is shown in Fig. 6 (a)- (c)[88], where (i) a homemade dimethylformide (DMF) solution, (ii) a homemade dichloroethane (DCE) solution, and (iii) a commercial catalytic particle-free aqueous SWNT solution (from Brewer Science Inc[83]) were used. It was found that bundles of SWNT were assembled for both the DMF and DCE solution and catalytic particle were attached to the SWNTs whereas the commercial surfactant free solution gives individual SWNT free from any catalytic particle. Along with being catalytic particle free, the SWNTs solution also needs to be highly stable over longer period of time which allows reproducible assembly.

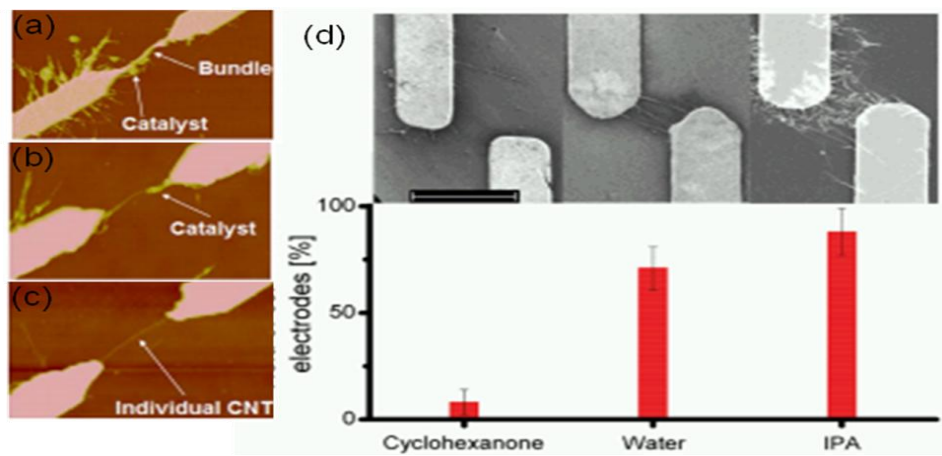

**Fig. 6.** Effect of the presence of catalyst particle on DEP assembly of individual SWNT (a) from DMF solution, (b) from DCE solution, and (c) from Brewer Science solution. Separation between the source and drain electrode is 1µm. Reproduced with permission from [85], P. Stokes et al., *J. Vac. Sci. Technol. B* 28, C6B7 (**2010**). © 2010, American Institute of Physics. (d) Statistics of electrode connection yield after dielectrophoresis.of CNTs dispersed in cyclohexanone, water, and IPA Representative SEM micrographs are presented for each solvent with a scale bar of 5 µm. Reproduced with permission from [100], M. Duchmap et al., *ACS Nano* 4, 279 (**2010**). © 2010, American Chemical Society.

Properties of the solvent in which SWNTs are immersed also play an important role in the DEP assembly. Since the conductivity, permittivity, and dielectric constant of SWNTs and various solvents are different,[56, 59, 100, 110-114] the DEP force is therefore unique for each solvent used, resulting in differences in the assembly yield. SWNTs immersed in a solvent with a low conductivity experience a higher DEP force compared to a solvent with high conductivity.[56] Besides, solvent with a low dielectric permittivity does not contribute to the DEP force resulting in low DEP yield.[56, 100] For example, the dielectric permittivity of both DI water and IPA is higher compared to that of cyclohexanone, which results in a higher assembly yield as shown in Fig. 6 (d).[100] In addition, volatile solvents represent additional challenges as the solvent can be evaporated before the assembly can be completed.

The DEP force, and hence the assembly can be manipulated significantly by varying the magnitude and frequency of the applied AC voltage. At low voltage the assembly yield is low as the DEP force is not strong enough for the SWNTs to align and assemble. The yield increases with the increase of applied voltage[98, 115]. However, application of a very high voltage can break the assembled SWNTs and therefore reduces the assembly yield. Typically a peak to peak voltage ($V_{pp}$) between 2V/µm to 5V/µm has been used for optimal yield.

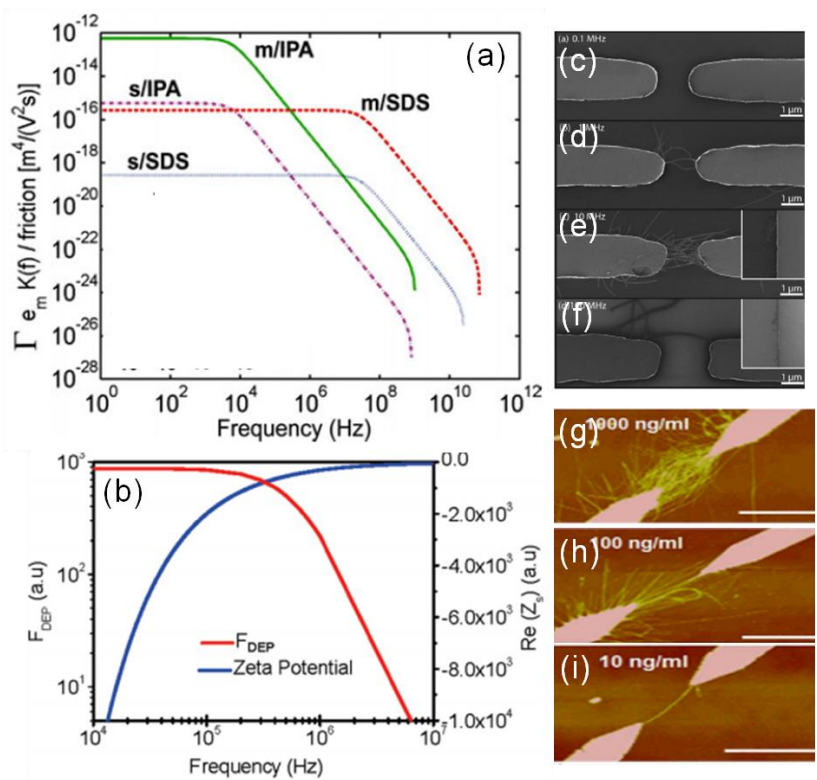

**Fig. 7.** (a) A plot of real part of Clausius-Mossotti (CM) factor as a function of frequency. Reproduced with permission from [56], M. Dimaki et al., *Nanotechnology* 15, 1095 (**2004**). © 2004, IOP Publishing Ltd. (b) Plot of DEP force (left axis) and zeta- potential (right axis) as a function of frequency. Reproduced with permission from [47], B. K. Sarker et al., *ACS Nano* 5, 6297 (**2011**) © 2011, American chemical Society. Frequency dependent DEP deposition yield studies at constant potentials. (c) 100 kHz (d) 1 MHz. (e) 10 MHz, and (f) 100 MHz. Scale bar: 1µm. Reproduced with permission from [103], B. R. Brug et al., *J. Appl. Phys.* 107, 124308 (**2010**). © 2010, American Institute of Physics. Dependence of solution concentration on the assembly of SWNT (g) 1000 ng/ml (h) 100 ng/ml, and (i)10 ng/ml in the solution. Scale bar: 1µm . Reproduced with permission from [85], P. Stokes et al., *J. Vac. Sci. Technol. B* 28, C6B7 (**2010**). © 2010, American Institute of Physics.

Frequency of the applied AC voltage plays a crucial role during DEP assembly of SWNTs.[59, 101, 103, 108] From DEP equation it is seen that the DEP force depends on the Clausius-Mossotti (CM) factor, which depends on the frequency. Dependence of real part of CM factor as a function of frequency for both metallic and semiconducting SWNTs is shown in Fig. 7 (a).[56] It can be seen here that for a particular solvent at a particular frequency the CM factor is higher for m-SWNTs than s-SWNTs and therefore the m-SWNTs experience greater DEP force. It can also be seen that the DEP force is higher at low frequency and is smaller at very high frequency. However, at low frequency the DEP assembly is strongly opposed by the surface charge induced zeta-potential (Fig. 7(b)) giving low assembly yield.[47, 116, 117] At high frequency the zeta-potential diminishes and a higher DEP assembly yield is obtained. Highest assembly yield for individual SWNTs are typically observed at frequency between 1-10 MHz.[38-40] The effect of frequency on the DEP deposition yield is shown in Fig. 7 (c) – (f).[103]

DEP assembly of the SWNTs greatly depends on the concentration of SWNTs in the solution. By varying the concentration of SWNTs in the suspending medium, it is possible to vary the number of SWNTs in the channel. Figure 7(g)-(h) presents the effect of solution concentration on the DEP assembly of SWNT.[85] In this case, the applied voltage, the frequency and the assembly time was kept fixed. A lot of SWNT assembled in the channel when a high solution concentration is used. By systematically reducing the solution concentration, individual SWNTs can be assembled in the channel.[85, 109] It is important to note that the solution concentration and DEP time are interrelated.[118] The density of SWNTs in the channel can also be varied to some extend by varying the DEP time while keeping the solution concentration fixed. For a particular solution concentration the SWNT density increases with the increase of DEP time. By keeping one of them fixed and varying the other parameter it is possible to control the number of SWNTs in the channel.[92, 119]

The dielectric oxide thickness and substrate conductivity also play an important role for the DEP assembly of SWNT.[100, 102] In particular, during simultaneous deposition method, the SWNT assembly yield is greatly influenced by the $SiO_2$ thickness.[102] For a particular areal size of the electrodes, the capacitance between the drain electrode and gate electrode is inversely proportional to the thickness of $SiO_2$. By optimizing the gate oxide thickness it is possible to obtain an optimum capacitive coupling which can provide a high deposition yield. For example, using a 800nm $SiO_2$ layer, it was shown by Vijayaraghavan et al. that an assembly yield of up to 90%[38] is possible whereas 250 nm $SiO_2$ gives only ~ 25%.[40, 92, 94]

For a particular type of SWNTs solution and electrode pattern the DEP parameters can be optimized for high yield assembly of individual SWNTs. Changing the SWNTs solution or electrode geometry requires re-optimization of DEP parameters for the new design. Since different groups use different kinds of SWNT solutions, surfactants, substrates and electrode patterns, the optimized parameters are not universal and are different for each group. However, once the parameters are optimized for a particular design, it will consistently give a similar device yield.

# 6. ELECTRONIC TRANSPORT PROPERTIES OF DEP ASSSEMBLED INDIVIDUAL SWNTS

We have already discussed that DEP offers a promising technique for the scalable and parallel fabrication of individual SWNT for device application. However, a common question is that whether solution processing techniques introduces defects in the SWNTs and degrades the intrinsic electrical properties of SWNTs which can limit their application in nanoelectronic devices? The properties of the DEP assembled devices can be evaluated from the electron transport measurements and several research groups have directed their research efforts towards that.[38-40, 85, 91-94, 120] In this section, we will present an overview of the electrical transport properties of the DEP assembled individual SWNT devices.

## 6.1 Properties of individual SWNT devices assembled from mixed SWNT solution

When as synthesized SWNTs are used for making stable solution, it contains both metallic and semiconducting SWNTs (mixed SWNTs solution). As a result, the DEP assembled individual SWNTs devices also contain either metallic or semiconducting SWNT in the channel. Typically, the mixed solution has about 2/3 s-SWNTs and 1/3 m-SWNTs.[121, 122] However, after the DEP assembly it was found that more than 50% of the devices are metallic while the rest are semiconducting.[38-40, 78, 79] The higher yields for metallic devices are attributed to larger DEP force experienced by m-SWNTs. The device properties are typically measured in a three terminal geometry with highly doped Si/SiO$_2$ substrate working as a gate. Figure 8 (a) shows the drain current ($I$) plotted as a function of back gate voltage ($V_{BG}$) of a representative metallic and semiconducting device for a fixed source drain voltage ($V_{DS}$).[60] As expected, the metallic nanotubes (m-SWNTs) show a weak modulation or no modulation in $I$ as a function of $V_{BG}$, whereas semiconducting nanotubes (s-SWNTs) show several orders of magnitude change in $I$ as a function of $V_{BG}$.

High quality contact is one of the major requirements for nanoscale device fabrication. It was found that the two-terminal contact resistances of as assembled SWNTs devices shows a wide ranges variation from a few KΩ to a few hundred MΩ. The high contact resistance has

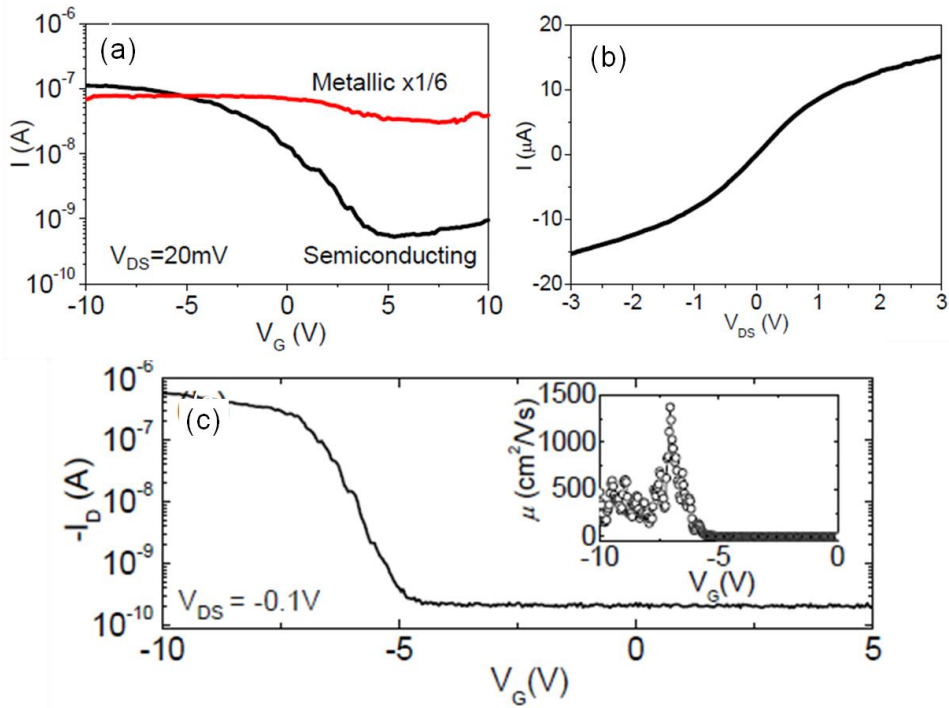

**Fig. 8** (a) Drain current ($I$) versus gate voltage ($V_G$) characteristics for semiconducting and metallic devices. (b) High bias $I$-$V_{DS}$ characteristic for a low contact resistance device showing saturation current of ~15 µA. Reproduced with permission from [85], P. Stokes et al., *J. Vac. Sci. Technol. B* **28**, C6B7 (**2010**). © 2010, American Institute of Physics. (c) Drain current ($I_D$) vs. gate voltage ($V_G$) at $V_{DS}$ = -0.1V for s-SWNT FET device (with $d$~1.3 nm). Inset: Mobility versus gate voltage showing a peak mobility value of ~1380 cm$^2$/Vs. Reproduced with permission from [40], P. Stokes et al., *Appl. Phys. Lett.* **96**, 083110 (**2010**). © 2010, American Institute of Physics.

been attributed to the presence of residual surfactant on the SWNT surface which causes poor contact at metal-SWNTs interface.[85, 91, 123] Annealing the SWNTs devices at high temperature (200-250°C) for 1 to 2 hour reduces the contact resistance by 1-2 orders of magnitude by removing the surfactants from the surface of the SWNTs[38-40]. Figure 8(b) shows a plot of $I$ versus $V_{DS}$ up to 3 V for a high quality m-SWNT device giving a saturation current of ~15 µA[85]. This is similar to a CVD grown m-SWNT with Ohmic contact, demonstrating that a high quality contact can be achieved from DEP assembled metallic devices.[124]

s-SWNTs devices also show large variation in device performance such as current on-off ratio and mobility. However, high mobility DEP assembled devices has also been demonstrated whose performance is similar to high quality CVD grown s-SWNTs device. Figure 8(c) shows transfer characteristics (drain current, $I_D$ plotted as a function of $V_{BG}$) of a high quality DEP assembled individual s-SWNTs device at $V_{DS}$ = -0.1 V.[40] The device shows p-type FET behavior with a high current on/off ratio (~ $3\times10^4$), a high on state conductance ($G_{on}=I_{on}/V_{DS}$) ~ 6 µS together with a transconductance of 1.15 µS which are close to the performance achieved by high quality back gated CVD grown SWNT devices.[125] Mobility, of the device was calculated from the transfer characteristics curve using the relation $\mu= (L^2/C_G\times V_{DS}) (dI_{DS}/dV_G)$, where $L$ is the channel length and $C_G= (2\pi\varepsilon L)/ln(1+2t_{ox}/r)$ is the gate capacitance, where $\varepsilon$ ~$3.9\varepsilon_0$ is the effective dielectric constant of $SiO_2$, $h$ is the thickness of the oxide, and $r$ is the radius of the s-SWNTs.[124] The device (with $r$ = 0.65 nm) gives a peak mobility of $\mu$ ~ 1380 cm$^2$/Vs (inset of Fig. 8(c)), which is close to the value obtained from theoretical approximation[125, 126] suggesting that high quality, defect free SWNTs devices can be fabricated using DEP.

The use of mixed solution for assembly possesses a severe limitation in transistor yield. Since m-SWNTs feels a greater DEP force then s-SWNTs, more than half of the assembled SWNTs are metallic when mixed SWNT solution is used for the DEP assembly. This reduces the overall FET yield, as only s-SWNTs can function as FET. This problem can be solved by using all s-SWNT solution during DEP assembly. In the next subsection we will discuss the electron transport properties of DEP assembled devices from semiconducting rich SWNTs solution.

**6.2 Properties of individual SWNT devices assembled from semiconducting rich SWNT solution**

Recently, solution based sorting techniques have been developed to separate SWNTs by their electronic type (chirality) [37, 88-90] which opens up an opportunity for the DEP assembly of semiconducting rich SWNT in order to obtain higher FET yield. Krupke group and Khondaker group independently studied the DEP assembly of s-SWNTs using s-SWNTs aqueous solution of 99% purity obtained by the density gradient ultracentrifugation technique.[91, 92, 94] It was found that the assembly yield for individual s-SWNTs is 20-30%. Figure 9(a) show transfer characteristics of such an individual s-SWNTs device.[92] The device shows p-type FET behavior with a high current on/off ratio ($I_{on}/I_{off}$) ~ $10^5$. The current on/off ratio ($I_{on}/I_{off}$) histogram of 63 such individual s-SWNT devices are shown in Fig. 9(b).[92] The most important point of this figure is that about 97% of the devices show $I_{on}/I_{off}$ >10 [39, 94] implying a 97% FET yield. However, a device-to-device variation in the FET performance is also observed for those devices. Such variations may be explained by the differences of chirality and diameter from tube

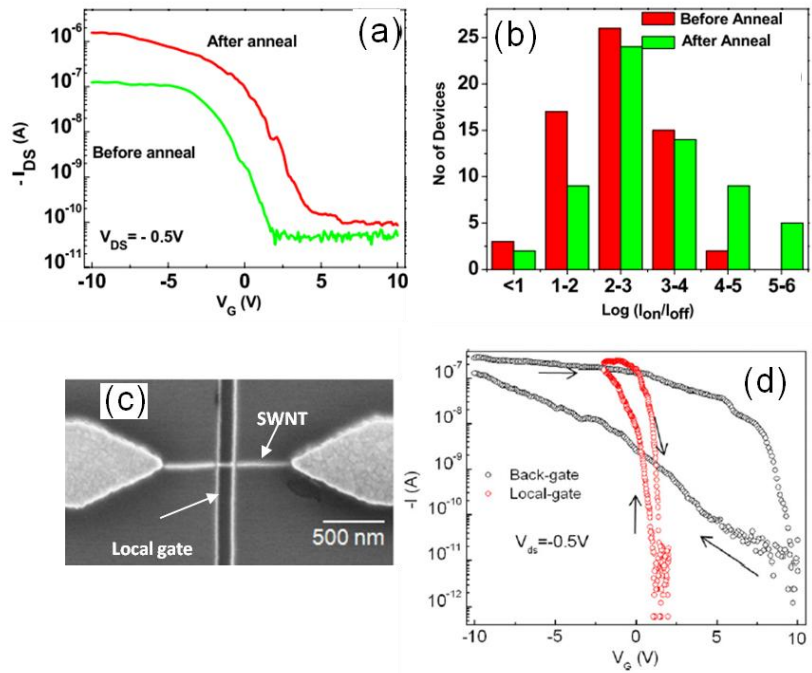

**Fig. 9** (a) Transfer characteristics of a representative individual SWNT FET device before and after annealing. (b) Histogram of $I_{on}/I_{off}$ for individual s-SWNT devices using all semiconducting SWNT solution before and after annealing. Reproduced with permission from [92], M. Islam et al., *Nanotechnology* 23, 125201 (**2012**). © 2012, IOP Publishing Ltd. (c) SEM image of a local gated device based on an individual s-SWNT. (d) Comparison of back-gated and local-gated transfer characteristics. The local gate shows much faster switching behavior and reduced hysteresis compared to the back gate. Reproduced with permission from [94], K. J. Kormondy et al., *Nanotechnology* 22, 415201 (**2011**). © 2011, IOP Publishing Ltd.

to tube which affect the band-gap and contact resistance of the s-SWNT, and in turn the mobility and on/off current ratio of the devices.[95, 125, 127, 128]

Figure 9(a) also show that it takes about 5 V to switch from on state to off state giving a subthreshold swing (S) of 1200 mV/dec. The high subthreshold swing is attributed to the thick $SiO_2$ gate oxide (250 nm). However, small S (~60mV/dec) are preferred in FETs for low power consumption and high speed operation[7]. Most of the DEP assembled back gated devices have S ~1000-2000 mV/dec,[91-94] which are too large for logic operation and far from their maximum capability. For SWNT FET the subthreshold swing is given by, $S=(2.303KT/q)(1+C_{IT}/C_G)$.[129, 130] Where, $C_G$ is the gate oxide capacitance and $C_{IT}$ is the capacitance due to interface trap charges. The subthreshold swing along with the hysteresis can be improved by employing local gated device geometry with a thin gate oxide. Khondaker group demonstrated a local gated geometry[94, 120] where individual SWNT were assembled on a $Al/Al_2O_3$ local bottom gate located at the middle of the channel between the source and drain electrodes, shown in Fig. 9(c).[107] Figure 9(d) presents a comparison of transfer characteristics between the local-gate and back-gate for a s-SWNTs device.[94] The back-gated device shows a large S~1000 mV/dec with huge hysteresis of ~ 7V. On the other hand, the local gate device shows a significantly reduced S ~

200 mV/dec and hysteresis of ~ 1 V. S values as low as 140 mV/dec had been reported, which is comparable to those for the CVD-grown $Al_2O_3$ gated devices.[131, 132] The faster switching behavior with the local gated device can be mainly attributed to the much thinner $Al_2O_3$ layer (~2–3 nm) compared to the thick layer of $SiO_2$ (250 nm). Besides, the reduced hysteresis observed in local gated device is a consequence of less interface traps at the SWNT/local gate oxide interface[133-135] which also helps reducing S.

## 6.3 Towards high yield fabrication of all SWNT FET devices via DEP

Even though 97% of the devices show FET when high purity s-SWNTs solution was used, the low assembly yield for individual s-SWNTs (~20-30%) reduces the overall device fabrication yield. The DEP assembly yield can be increased to 90% if the assembly condition is optimized such that in addition to individual SWNT, 2-5 SWNTs are allowed to connect per electrode pair (Fig 10 (a)). Electron transport study shows that the devices having 2-5 s-SWNTs also shows good FET behavior. For example, Figure 10(b) shows the transport characteristics of a device with 2-SWNTs in the channel.[92] It gives p-type FET behavior with a current on-off ratio ~ $2\times10^4$ and an on-conductance of 3.4 μS with a mobility of 126 $cm^2$/Vs. The mobility was calculated using the formula $\mu= (L/WC_iV_{DS})(dI_{DS}/dV_G)$, with $C_i=D/[C_Q^{-1}+(1/2\pi\varepsilon)\ln[\sinh(2\pi t_{ox}D)/\pi Dr]]$, where D is the linear density of the SWNT, W is the channel width, and $C_i$ is the specific capacitance per unit area of aligned array with $C_Q$ (=$4\times10^{-10}$ $Fm^{-1}$) [45]. For this calculation L=1 μm was used while an upper limit of W=0.25 μm for 2 SWNT was determined from SEM image giving D=8.[80] Statistics for the $I_{on}/I_{off}$ of the devices made from 1-5 SWNTs are shown in Fig. 10(c).[92] It shows that 90% of such devices show FET behavior. Using such a scheme, an assembly yield of 90%[80, 86] with an overall FET yield of 90% had been demonstrated.[92] This is in contrast to the mixed s-SWNT solution where the assembly of more than one SWNT per site significantly increases the chance of getting at least one m-SWNTs per site resulting in metallic behavior of the device. A FET yield of 90% with an assembly yield of 90% obtained by DEP is by far the best for any type of device assembly technique[92].

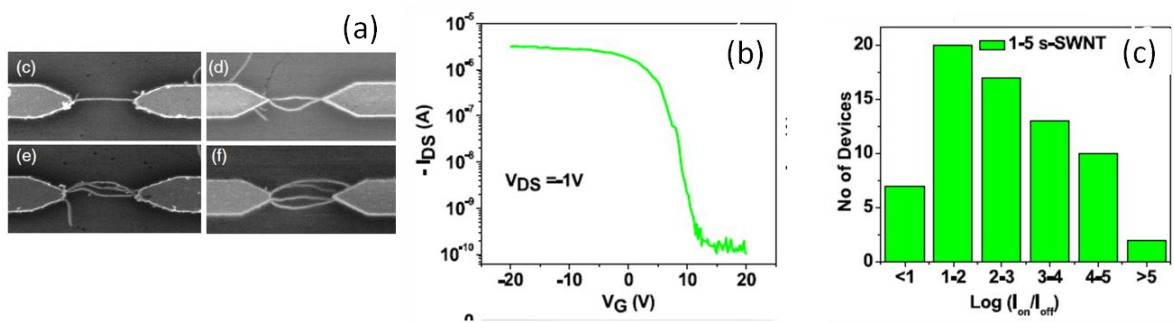

**Fig. 10:** Local view of the SEM images of the devices containing (a) one, two, three, and four s-SWNTs between the electrodes. Separation between the source and drain electrode is 1μm. (b) Transfer characteristics of a FET device with two s-SWNTs. (c) Histogram of the $I_{on}/I_{off}$ for all devices with one to five s-SWNTs. Reproduced with permission from [92], M. Islam et al., *Nanotechnology* 23, 125201 (**2012**). © 2012, IOP Publishing Ltd.

## 7. CONCLUSION AND FUTURE OUTLOOK

In conclusion, we presented an overview on the DEP assembly of individual SWNTs towards the realization of scaled up fabrication of SWNTs devices for practical applications. We have discussed the theory of DEP, presented a historic overview with an emphasis on the recent progress on the DEP assembly of individual SWNTs devices, and discussed the electronic transport properties of the fabricated devices. From these discussions, it is clear that the DEP assembly of SWNTs has made tremendous progress during the last two decades. With continuous progress in solution processing, it is now possible to obtain high quality and catalytic particle free stable SWNTs solution that are chirally sorted. High performance DEP assembled SWNT devices has also been demonstrated. With semiconducting rich SWNTs, a 90% transistor yield with 90% assembly yield has become a reality.

While significant progress has been made, there still exists a number of scientific and technical challenges that need to be addressed for wide scale application of the SWNT based devices. One of them is a large device-to-device performance variation (contact resistance, mobility, on-off ratio) in the DEP assembled SWNT devices. Even though the technique has been developed to separate metallic and semiconducting SWNTs, the wide range of variation in the diameter of SWNT, causes individual SWNTs based devices to show variation in their performance such as mobility and current on-off ratio. Very recently, effective technique had been developed demonstrating narrow distribution of diameters and chirality of SWNTs[136, 137] which could facilitate more homogeneous device performance and the future effort should be directed in the DEP assembly of uniformly mono-dispersed SWNTs for device application. In addition, techniques need to be developed to improve the contact resistance of the DEP assembled SWNTs. Contact resistance also depends on the diameter of the nanotube as well as the surfactant used for the processing. Improving the diameter distribution will also help reduce the contact resistance inhomogeneity. Large-scale assembly of ultra short (sub-100 nm) SWNTs has not been demonstrated yet, which is necessary for device scaling. Besides, complex device structures such as inverter or logic circuits also need to be demonstrated using DEP. Addressing these challenges along with the progress already made will allow parallel fabrication of CMOS compatible individual SWNT devices for future nanoelectronic application.

**Acknowledgement:** The authors acknowledge financial support from U.S. National Science Foundation under Grant No. ECCS-0748091 (CAREER)